\documentclass[12pt]{article}
\pdfoutput=1 
\usepackage{jheppub}
\usepackage{amsmath}
\usepackage{amssymb}
\usepackage{graphicx}
\usepackage{mathrsfs}
\usepackage{subcaption}
\usepackage{xcolor}
\usepackage[utf8]{inputenc}

\usepackage{subcaption}
\usepackage{float}
\usepackage{afterpage}
\renewcommand{\figurename}{Fig.}

\newcount\xrefpos \xrefpos=0
\newcount\yrefpos \yrefpos=0
\newcount\xput \xput=0
\newcount\yput \yput=0

\def\refpos#1 #2 #3{\global\xrefpos=#1 \global\yrefpos=#2
                         \rlap{$\smash{#3}$}}
\def\put #1 #2 #3{\xput=#1 \yput=#2
                  \advance\xput by -\xrefpos
                  \advance\yput by -\yrefpos
                  \rlap{\kern\the\xput truebp
                        \vbox to 0pt{\vss\hbox{$\displaystyle #3$}
                        \kern\the\yput truebp}}}
\def\beginlabels\refpos#1\endlabels{\hbox{$\refpos#1$}}

\newcommand{\be}{\begin{equation}}
\newcommand{\ee}{\end{equation}}
\newcommand{\bea}{\begin{eqnarray}}
\newcommand{\eea}{\end{eqnarray}}
\def\bse{\begin{subequations}}
\def\ese{\end{subequations}}

\def\IZ{\relax\ifmmode\hbox{Z\kern-.4em Z}\else{Z\kern-.4em Z}\fi}

\def\R{\mathrm{R}}

\definecolor{rust}{rgb}{0.8,0.2,0.2}

\title{On Brane Instabilities in the Large $D$ Limit}

\author{\! Moshe Rozali}
\author{\!, Alexandre Vincart-Emard}
\affiliation{
Department of Physics and Astronomy, University of British Columbia,\\
Vancover, BC V6T 1Z1, Canada}
\emailAdd{rozali@phas.ubc.ca}
\emailAdd{ave@phas.ubc.ca}
\abstract{
Using an expansion in large number of dimensions, taken to subleading orders, we discuss several issues concerning the Gregory-Laflamme instabilities. We map out the phase diagram of neutral and charged black strings, and comment on the possible transition in the nature of the final state of the instability at higher order in the $1/D$ expansion. We also discuss unstable black membranes, and show that in certain limits the preferred shape of the non-uniform phase is a triangular lattice.}

\begin{document}
\maketitle

\section{Introduction}
\label{sec:intro}

Since its discovery, the Gregory-Laflamme instability \cite{GL1993,GL1994} has been a source of  many  insights into General Relativity and its extended black brane solutions in higher dimensions. The fate of the instability for a string is much studied (for a comprehensive review, see \cite{Obers:review}): there is now strong evidence that the end-point of the black string instability depends on the number of spacetime dimensions. It was shown in \cite{Sorkin2004} that there exists a critical dimension $D^* = 13.5$ above which non-uniform black strings (NUBS) become stable and have larger horizon areas than their uniform counterparts, thus making them natural candidates as the end-point of the GL instability. Below $D^*$, numerical simulations \cite{Choptuik2003,Lehner2011} have presented evidence that black string horizons bifurcate in a self-similar cascade of black holes pinching off to arbitrary small scales along the string direction, thus violating the cosmic censorship hypothesis (despite the arguments proposed by \cite{Horowitz2001}). A numerical evolution beyond the critical dimension would be a welcome addition, but the high-performance computing resources required would be an obstacle to this endeavour. \\
\\
A different approach --- general relativity in the limit where the number of dimensions is large \cite{LargeD} --- offers a promising framework in which one can address such questions analytically, or numerically with only modest resources. Despite the theory being formally valid only when $D \rightarrow \infty$, its potential even at finite $D$ was highlighted in a range of applications, ranging from the striking agreement of large $D$ black holes quasinormal modes for both large and small values of $D$, to the alternative derivation of the critical dimension $D^*$ found in \cite{CriticalD}. \\
\\
In this paper we discuss different aspects of the phase structure of the non-uniform black objects. Following in the footsteps of \cite{EmparanBS,EmparanCharged}, we perform our analysis by promoting the mass, charge and momentum densities on the string to be collective variables, and solve the resulting equations numerically, observing their conserved charges at asymptotic infinity. 
Using this approach,  we discuss several issues concerning the end-state of the Gregory-Laflamme instability of extended black objects.\\
\\
Discussing the general charged black string, we find that the entropy difference between the non-uniform configuration and its uniform counterpart remains finite and positive for all such charged black strings, even in the extremal limit. Indeed, in the extremal limit we are able to show that fact analytically. Thus we conclude that there is a second order transition to a non-uniform phase for all charged NUBS, which are entropically favoured despite the weakening of the GL instability due to the addition of electric charge. \\
\\
We also investigate the physics of the neutral string to next-to-leading order (NLO). Our goal is to find the signal, in the large $D$ expansion, of the transition of the instability end-point from a non-uniform black string to a pinch-off scenario. Indeed, below the critical dimension $D^*$  where the NUBS have lower entropy than the uniform string, there is a different end-point to the instability, which is expected to be a pinch-off. While we find signs that this is indeed the case, we are unable to find a universal value for the associated critical dimension (which may be different from $D^*$) from our analysis.\\
\\
Lastly, we turn our attention to the phase structure of two-dimensional unstable membranes on oblique lattices. By comparing brane solutions of different shapes, we find that the triangular lattice configuration is the one that minimizes the corresponding thermodynamic potential for localized 2-branes.\\
\\
The outline of the paper is as follows: In Section \ref{sec:setup}, we summarize how to obtain charged (and neutral) black string solutions in the characteristic formulation of general relativity, at leading order in $D$. This serves to set up our notations and explains our numerical method. In Section \ref{sec:phaseNUBS}, we discuss the phase structure of charged black string. To that end, we find the subleading corrections to the metric and gauge field, necessary to discuss the entropy difference (in the micro-canonical ensemble) between the uniform and non-uniform solutions. We find that a charged non-uniform string always have a larger horizon area than uniform configurations, even in the extremal limit.  We also obtain the equations governing the dynamics of the $1/D$ corrections to the mass and momentum densities, and discuss stability conditions of the neutral string to next-to-leading order, and the signals that there may be a transition to pinch-off as the final state as we lower $D$. Lastly, we explore the thermodynamics of unstable two-dimensional branes on general oblique lattices in Section \ref{sec:lattice}. We find that the preferred shape of the lattice is triangular, up to small deviations, likely due to finite size effects.

%



\section{Charged $p$-brane Solutions}
\label{sec:setup}

We start our analysis by finding static charged black string solutions. Keeping in mind that we are interested in finding non-uniform black string solutions, we must allow redistribution of mass and charge to occur along the spatial direction. To this end, we introduce a local Galilean boost velocity and promote the mass and charge densities to vary along the string. We then solve the Einstein-Maxwell equations at leading and subleading orders, from which we extract the effective equations that describe the nonlinear fluctuations of the string horizon. 
\subsection{Uniformly Charged $p$-Branes}
\label{sec:uniform}

The equations of motion that charged, spherically symmetric $p$-branes satisfy in the limit where $D \rightarrow \infty$ can be obtained by considering the Einstein-Maxwell action in $D = n + p + 3$ dimensions,
\be I_\text{EM} = \int d^D x \;\sqrt{G} \left( R_G - \frac{F^2}{4} \right), \label{EMaction} \ee 
where $F = dV$ is a Maxwell potential. Performing dimensional reduction on (\ref{EMaction}) so that the metric becomes of the form
\be ds^2_G = g_{\mu \nu}(x) dx^\mu dx^\nu + e^{\phi(x)} d \Omega_{n+1}^2, \ee
where the coordinates $x^\mu = (t, r, z^A)$ span a $p+2$ dimensional space, we obtain the action 
\be I_\text{EM} = \int d^{p+2} x \;\sqrt{g} \; e^{\frac{(n+1) \phi}{2}} \left( R_g + n(n+1) e^{-\phi} + \frac{n(n+1)}{4} ( \nabla \phi)^2 - \frac{F^2}{4} \right). \label{action} \ee
The equations of motion that follow from (\ref{action}) are \cite{MembraneQ}
\begin{align} \begin{split} \mathbb{E}_{\mu \nu} =  R_{\mu \nu} - \frac{n+1}{2} \nabla_\mu \nabla_\nu \phi - \frac{n+1}{4} \nabla_\mu \phi \nabla_\nu \phi - \frac{1}{2} \left( F_{\mu \alpha} F_\nu^{\; \; \alpha} - \frac{1}{2(n+p+1)} F^2 g_{\mu \nu} \right) =& \; 0 \\
\mathbb{J}^\mu = \nabla_\alpha F^{\alpha \mu} + \frac{n+1}{2} \nabla_\alpha \phi \; F^{\alpha \mu} =&\; 0 \\ 
n \;e^{-\phi}  - \frac{n+1}{4} (\nabla \phi)^2 - \frac{1}{2} \nabla^2 \phi + \frac{1}{4(n+p+1)}F^2 =& \; 0. \label{EMeqns}
 \end{split} \end{align}
The solution to these equations is known \cite{HigherBH}: non-dilatonic black $p$-branes in the presence of an electric potential have the metric
\be ds^2 = -\frac{f}{h^2} dt^2  + h^B \left( \frac{dr^2}{f} + r^2 d\Omega_{n+1}^2 + d\vec{z}^{\;2} \right), \ee
where
\be f(r) = 1 - \left( \frac{r_0}{r}\right)^n, \;\;\;\;\; h(r) = 1 + \left( \frac{r_0}{r}\right)^n \sinh^2 \alpha, \;\;\;\;\; B =  \frac{2}{n+p}, \ee
whereas the gauge field $V$ has solution
\be V = - \frac{\sqrt{N}}{h} \left( \frac{r_0}{r}\right)^n \sinh \alpha \cosh \alpha \; dt, \;\;\;\;\; N \equiv B + 2. \ee
In these coordinates, the outer horizon is located at $r = r_0$, whereas the inner horizon coincides with the singularity at $r=0$. 

\subsection{Characteristic Formulation for a Charged Black String}
\label{sec:characteristic}

To describe a non-uniform charged black string,  we start by a more general ansatz where the black string is locally boosted along its worldvolume $z^a = (t,z^A)$. Doing so is easier in the characteristic formulation of general relativity, where the metric is expressed in terms of the ingoing Eddington-Finkelstein (EF) coordinates. For a general Lorentz boost $u^a$, the EF coordinates $\sigma^a = (v, x^A)$ take the form \cite{Camps2010,AFRN}
\be \sigma^a = z^a + u^a r_*, \;\;\;\;\; r_*(r) = r +  \int_r^\infty \frac{f - h^{N/2}}{f} dr. \ee
The boosted metric for the charged string becomes
\be ds^2 = h^B \left( -\frac{f}{h^N} u_a u_b d\sigma^a d\sigma^b -2 h^{-N/2} u_a d\sigma^a dr + \Delta_{a b} d\sigma^a d\sigma^b + r^2 d\Omega^2_{n+1}  \right), \ee
where $ \Delta_{ab} = \eta_{ab} + u_a u_b $ is the orthogonal projector defined by the boost vector. Similarly, the gauge potential $V$ becomes
\be V = - \frac{\sqrt{N}}{h} \left( \frac{r_0}{r}\right)^n \sinh \alpha \cosh \alpha \; u_a d\sigma^a, \ee
and we take the radial gauge in order to set $V_r = 0$. \\
\\
Our aim is to find solutions to the Einstein-Maxwell equations in which the black string's energy and charge densities, as well as the boost velocity along the string, are promoted to collective coordinates that vary in time along the $x$-direction. Given the hierarchy of scales present in the large $D$ limit of black holes, we must specify the length scale relevant to the physics we wish to explore. It is known that black branes are unstable when subjected to perturbations of wavelength $\sim r_0/\sqrt{D}$. As such, we need to rescale the direction $x$ along the string, $dx \rightarrow dx/\sqrt{n}$, thus making the boost non-relativistic. Additionally, the quasinormal modes under consideration scale like $\omega \sim \mathcal{O}(D^0)$, implying that the dynamics of the near-horizon geometry is decoupled from the asymptotic region \cite{EmparanQNM}. Consequently, we will require the metric components to be asymptotically flat and the potentials to vanish at infinity at all orders in the perturbative expansion. \\
\\
We write the metric and the gauge potential in terms of unknown fields
\be ds^2 = - A dv^2 + 2 u_v dv dr + 2 u_a dx^a dr - 2 C_a  dx^a  dv+ G_{ab} dx^a dx^b,  \label{metricEF} \ee
\be V = V_0  \; dv + V_a \; dx^a, \label{gaugeEF} \ee
for which we allow the following $1/n$ expansion:
\be A =\; \sum_{k \geq 0} \frac{A^{(k)}(v,x,\R)}{n^k}, \;\;\; u_v =\;  \sum_{k \geq 0} \frac{u_v^{(k)}(v,x,\R)}{n^k}, \;\;\;C_a =\;  \sum_{k \geq 0} \frac{C_a^{(k)}(v,x,\R)}{n^{k+1}}, \ee
\be G_{ab} = \; \frac{1}{n} \left( 1 +\sum_{k \geq 0} \frac{G_{ab}^{(k)}(v,x,\R)}{n^{k+1}} \right), \;\;\; u_a = \; \frac{u_a^{(0)}}{n} + \sum_{k \geq 1} \frac{u_a^{(k)}(v,x,\R)}{n^{k+1}}, \ee 
\be V_0 =\;  \sum_{k \geq 0} \frac{V_0^{(k)}(v,x,\R)}{n^k}, \;\;\;V_a =\;  \sum_{k \geq 0} \frac{V_a^{(k)}(v,x,\R)}{n^{k+1}}, \ee
where the new radial coordinate $\R = (r/r_0)^n$ is well-suited for near-horizon analysis. For the scalar field, we make the choice to keep $\phi = \log\left( r^2 h^B\right) $ at all orders in the expansion in order to maintain the spherical symmetry of the solution. Also note that demanding $u_a$ to be a constant at leading order is simply a gauge choice; we will set $u_a^{(0)} = 0$ to fix the rest frame of the black string. \\

\subsection{Solutions and Effective Brane Equations at Leading Order}
\label{sec:collectiveleading}

At leading order, the solutions to the Einstein-Maxwell equations are given by
\begin{align} A^{(0)} =&\; 1 - \frac{m}{\R} + \frac{q^2}{2\R^2}, \;\;\;\;\; C_a^{(0)} = \left( 1 - \frac{q^2}{2m \R} \right) \frac{p_a}{\R}, \;\;\;\;\; u_v^{(0)} = 1, \\
G_{ab}^{(0)} =&\;  \left(1 - \frac{q^2}{2m \R} \right) \frac{p_a p_b}{m \R} - \log\left(1 - \frac{\R_-}{\R} \right) \left( 2 \delta_{ab}  + \partial_a \frac{p_b}{m} + \partial_b \frac{p_a}{m} \right), \label{Gxx0} \\
V_0^{(0)} =&\; - \frac{q}{\R}, \;\;\;\;\; V_a^{(0)} = \frac{q p_a}{m \R}. \end{align}
Note that the radial coordinate appearing above has been shifted so that the outer and inner horizons of the charged black hole are now located at
\be \R_\pm = \frac{1}{2} \left( m \pm \sqrt{m^2 - 2 q^2} \right). \ee
The collective variables $m$ and $q$ are directly related to the energy and charge densities $\mathcal{M}$ and $\mathcal{Q}$ of uniform $p$-branes
\be \mathcal{M} = r_0^n (n+1+n N \sinh^2 \alpha), \;\;\; \mathcal{Q} = n \sqrt{N} r_0^n \sinh \alpha \cosh \alpha. \ee
The large $n$ expansion of these conserved quantities shows a correspondence between the old and new effective fields on the branes at leading order: 
\be m \equiv r_0^n \cosh 2 \alpha \;\;\;\;\;  \text{and} \;\;\;\;\; q \equiv \frac{r_0^n}{\sqrt{2}} \sinh 2 \alpha. \ee
As for $p$, it is related to the momentum density on the black brane; the gauge choice $u_a^{(0)}=0$ ensures that the total momentum on the brane vanishes.\\
\\
The equations that govern the dynamics of the collective variables for general $p$-branes are\footnote{In the large $n$ limit, $\partial_t = \partial_v$ and the dynamics of the collective variables take place in Schwarzschild time.} \cite{EmparanCharged}:
\begin{subequations}
\begin{align} &\; \partial_t m - \partial_i \partial^i m = -\partial_i p^i, \label{Mdot} \\
&\; \partial_t q - \partial_i \partial^i q = - \partial_i \left( \frac{p^i q}{m} \right), \label{Qdot} \\
&\; \partial_t p_i - \partial_j \partial^j p_i =  \partial_i \left( \R_+ - \R_- \right) - \partial^j \left[ \frac{p_i p_j}{m} + \R_-\left( \partial_i \frac{p_j}{m} + \partial_j \frac{p_i}{m} \right) \right] \label{pdot} .\end{align} \label{subeqns} 
\end{subequations} 
These equations, which describe the fluctuations of black branes on the compactified string directions $x^i$ in the large $D$ limit, can also be written as conservation equations $\partial_\mu \tau^{(0)\mu \nu} = 0$ for a quasilocal stress tensor at $\R \rightarrow \infty$, whose components are
\begin{align} \tau_{00}^{(0)} =&\; m, \;\;\;\;\;\; \tau_{0i}^{(0)} = \partial_i m - p_i, \\
\tau_{ij}^{(0)} =&\; \partial_i \partial_j m - \left(\R_+ -\R_-\right) \delta_{ij} + \frac{p_i p_j}{m} - \left(\partial_i p_j +\partial_j p_i \right)\left(1 - \frac{\R_-}{m} \right).  \label{Txx0}
\end{align}
These equations are very easy to solve numerically. In doing so we discover the stable end-point of the charged string instability, as well as the time-dependent process leading to that end-point. To discuss the thermodynamics, to which we turn next, we need to discuss the next order in the large $D$ expansion.

\vspace{10 pt}

\section{Phase Structure of the NUBS}
\label{sec:phaseNUBS}

Having set up our equations and non-uniform solutions, we now turn our attention to the phase structure of the charged black string, compactified along $x \in [-L/2,L/2]$. A proper analysis first requires us to examine the properties of NUBS to subleading order in the large $D$ expansion. 

\subsection{Solutions and Effective Brane Equations at Subleading Order}
\label{sec:collectiveleading}

At subleading order for the charged black string ($p=1$), we find the solutions
\begin{align} A^{(1)} =&\; - \frac{\delta m}{\R} + \frac{q \delta q}{\R^2} - \frac{q^2}{2 \R^2} + \log\left( 1- \frac{\R_-}{\R} \right) \left[ - \frac{m}{\R} + \frac{q^2}{2 \R^2} \right] \left( \frac{p}{m} \right)^\prime \nonumber \\ 
&\; - \log \R \left[ \frac{p^\prime}{\R} - \frac{q}{\R^2} \left( \frac{pq}{m} \right)^\prime \right] \\
V^{(1)} =&\; - \frac{\delta q}{\R} - \frac{q}{\R} \log \left( 1- \frac{\R_-}{\R} \right) \left( \frac{p}{m} \right)^\prime - \frac{\log \R}{\R} \left( \frac{pq}{m} \right)^\prime  \\
u_v^{(1)} =&\; \frac{p^2}{2m^2} \left[ - \frac{m}{\R} + \frac{q^2}{2 \R^2} \right] - \left( 1- \frac{\R_-}{\R} \right)^{-1} \frac{\R_-}{\R} \left(\frac{p}{m}\right)^\prime 
 \end{align}
where a prime denotes differentiation with respect to $x$. It is straightforward to verify that $\delta m(\sigma)$ and $\delta q(\sigma)$, which appear as integration constants, indeed correspond to the $1/n$ corrections to the mass and charge densities by computing the ADM mass $\mathcal{M}$ and the electric flux at infinity via
 \begin{align} \mathcal{M} =&\;  - \oint_{S^{n+1}_\infty} \nabla^\mu \xi^\nu_{(v)} dS_{\mu \nu} = \int \left( m(\sigma) + \frac{\delta m(\sigma)}{n} + \cdots \right) dx, \label{ADM} \\
\mathcal{Q} =&\;  \frac{1}{2} \oint_{S^{n+1}_\infty} F^{\mu \nu} dS_{\mu \nu} = \int \left( q(\sigma) + \frac{\delta q(\sigma)}{n} + \cdots \right) dx,  \label{flux} \end{align}
where $\xi^\mu_{(v)} = \delta^\mu_v$ is a timelike Killing vector, and the integration is performed over $S^{n+1}$ at spatial infinity. Some ambiguity remains when using (\ref{ADM}) and (\ref{flux}) to define the mass and charge density corrections since a shift in either quantity by the derivative of a periodic function results in identical ADM mass and electrix flux. We thus examine the multipole expansion of $g_{00}$ and $V_0$ about asymptotic infinity to identify $\delta m(\sigma)$ and $\delta q(\sigma)$ as the appropriate corrections. Let us remark that $\mathcal{M}$ and $\mathcal{Q}$ remain conserved at all orders in the $1/n$ expansion, and as such the corrections $\delta m(\sigma)$ and $\delta q(\sigma)$ have vanishing integrals over the string direction. \\
\\
The above solutions will be useful in Section \ref{sec:entropy}. For the remainder of this section, we will focus our efforts on the neutral case. The string's momentum correction is found in the $dv dx$ component of the metric
\be C_x^{(1)} = \frac{\delta p}{\R} + \frac{\log \R}{\R} \left( \frac{p^2}{m} \right)^\prime; \ee
it is a quantity associated with the asymptotic Killing vector $\xi^\mu_{(x)} = \delta^\mu_x$
\be \mathcal{P} = \oint_{S^{n+1}_\infty} \nabla^\mu \xi^\nu_{(x)} dS_{\mu \nu} = \int \left( p(\sigma) + \frac{\delta p(\sigma)}{n} + \cdots \right) dx . \ee
The equations that govern the dynamics of $\delta m$ and $\delta p$ are\footnote{The large $D$ equations at NLO were first obtained in \cite{Herzog:2016} for asymptotically AdS spacetimes, albeit in a gauge different from ours. We have confirmed that the equations we obtain agree with their AdS counterpart up to a redefinition of the momentum correction and a few sign flips.}
\begin{subequations}
\begin{align}  \partial_t \delta m \;-&\; \delta m^{\prime \prime} + \delta p^\prime = F_{\delta m}^\prime, \\
\partial_t \delta p \;-&\; \delta p^{\prime \prime} - \left[ \left(1 + \frac{p^2}{m^2} \right) \delta m - \frac{2 p}{m} \delta p \right]^\prime = F_{\delta p}^\prime,  \end{align}
\label{colleqnscorr} \end{subequations}
with the source functions $F_{\delta m}$ and $F_{\delta p}$ given by 
\begin{align} F_{\delta m} =&\;  p + \left( m + 2 p^{\prime} -\frac{3}{2} \frac{p^2}{m} \right)^\prime  ,\\
F_{\delta p} =&\;  F_0 \log m  + \frac{F_0}{2} - \frac{p^2}{m} -\frac{3}{2} \left( \frac{p^3}{m^2} \right)^\prime + \frac{4 p m^\prime}{m} - \frac{2 p^2 m^{\prime \prime}}{m^2} + \frac{4 p p^{\prime \prime}}{m} ,
\end{align}
and
\be F_0 =2 m \left[ 1 + \left( \frac{p}{m} \right)^\prime \right] \left( \frac{p}{m} \right)^\prime .\ee
As is the case at leading order, these equations can also be rewritten as correction terms for the asymptotic stress tensor $\tau_{ij}$:
\begin{align} \tau_{00}^{(1)} =&\; \delta m, \;\;\;\;\;\; \tau_{0x}^{(1)} =  \delta p - \delta m^\prime - F_{\delta m}, \\
\tau_{xx}^{(1)} =&\; - F_0 (\log m - 3) + m - \delta m \left( 1 + \frac{p^2}{m^2} \right) + 2 \delta p \frac{p}{m} + \left(\delta m + 4m + 4 p^\prime - 7 \frac{p^2}{m} \right)^{\prime \prime} \nonumber \\
-&\; 2 \left( \delta p + 3p - \frac{3}{2} \frac{p^3}{m^2} \right)^\prime. \end{align}
The left-hand sides of equations (\ref{colleqnscorr}) correspond to the differential operators one would find at order $1/n$ by letting $m \rightarrow m + \delta m/n$ and $p \rightarrow p + \delta p /n$ in the collective equations (\ref{subeqns}). However, the presence of the source terms breaks Galilean invariance. Moreover, whereas the leading order equations are invariant under a rescaling of the mass and momentum, the collective equations for the correction terms are not. This can be understood as a consequence of the dependence of the black string temperature on its mass density $m_0$ at NLO. Indeed, when the NUBS is stationary, one can calculate the surface gravity $\kappa$ via the relation
\be \kappa^2 = \sqrt{ - \frac{1}{2} \nabla^\mu \xi^\nu_{(v)} \nabla_\mu \xi_{(v)\nu} }, \ee
and evaluation at the Killing horizon is understood. Rescaling the surface gravity such that the temperature is of $\mathcal{O}(1)$ at leading order, we obtain
\be T = \frac{\kappa}{2 \pi n} = \frac{1}{4\pi} - \frac{1}{4 \pi n}\left( \log m - \frac{ (m^\prime)^2 }{2m^2} + \frac{m^{\prime \prime}}{m} \right) = T^{(0)} + \frac{T^{(1)}}{n}, \ee
\be \text{with} \;\;\;\;\; T^{(1)} = - \frac{1}{4 \pi} \left( \log m - \frac{m^{\prime 2}}{2 m^2} + \frac{m^{\prime \prime}}{m} \right). \ee
As a consequence of this, we find that the shapes of $\delta m$ and $\delta p$ depend on the additional parameter $m_0$. However, $m_0$ should not be regarded as an independent parameter of our solutions. The initial state of our system is uniquely characterized by $D$ and the ratio $L/r_0$, and as such mass and momentum profiles at different initial temperatures contain the same information packaged differently. It is easier to work in units where $r_0 = 1$ (for which $m_0 = 1$ also), but we can equivalently rescale all dimensionful quantities by an appropriate power of $r_0$ to obtain the same information. \\
\\
Let us now turn our attention to the next-to-leading order correction to the dispersion relation for the black string. Linearized perturbations around the uniform black string solution $m(x) = m_0 + \Delta m e^{ - i \omega t + i k x}$, with momentum $k = 2 \pi/L$ aligned along the string direction $x$, allow for a non-trivial solution only if the condition
\be \left\vert
\begin{array}{cc}
-k^2+ i \omega - \frac{k^2}{n} & - i k + \frac{i k(1-2k^2)}{n} \\
 i k & -k^2+ i \omega - \frac{k^2 (1 + 2 \log m_0)}{n} \\
\end{array}
\right\vert = 0 \ee
is satisfied. Letting $\Omega = - i \omega$, the dispersion relation for the black string reads
\be \Omega(k) = k - k^2 - \frac{k}{2n} \left( 1 + 2 k + 2 k \log m_0 - 2 k^2 \right) + \mathcal{O}(n^{-2}). \ee
This yields the corrected threshold mode $\Omega(k_\text{GL})=0$ to be
\be k_\text{GL} = 1 - \frac{1 + 2\log m_0}{2n}+ \mathcal{O}(n^{-2}).\ee
Rewriting the above equations as $\hat{\Omega} (\hat{k}) \equiv \Omega(k r_0) r_0$ so that they become dimensionless eliminates the dependence on the mass density $m_0$
\be \hat{\Omega}(\hat{k}) = \hat{k} - \hat{k}^2 - \frac{\hat{k}}{2n} \left( 1 + 2 \hat{k} - 2 \hat{k}^2 \right) + \mathcal{O}(n^{-2}), \;\;\; \hat{k}_\text{GL} = 1 - \frac{1}{2n}+ \mathcal{O}(n^{-2}), \ee
and thus we recover the expected result \cite{LargeD}, regarding the shift of the critical wavelength of the Gregory-Laflamme instability at subleading order.



\subsection{Charged Black String Phase Diagram}
\label{sec:entropy}

When $k < k_\text{GL}$, neutral NUBS always have a lower event horizon surface area than uniform string solutions. Since the addition of charge weakens the GL instability of black strings, it is natural to wonder if the NUBS remains entropically favoured, especially as we approach extremality. \\ 
\\
The entropy $\mathcal{S}$ of a black string is related to the area of its horizon
\be \mathcal{S} = \frac{\Omega_{(n+1)}}{4 G} \int_\text{horizon} \sqrt{g_{xx}} e^\frac{(n+1) \phi}{2} \; dx = \frac{\Omega_{(n+1)}}{4 G \sqrt{n}}\left( \mathcal{S}_0 + \frac{\mathcal{S}_1}{n} + \cdots \right). \ee
At leading order, the area of a boosted string is given by the integral of the outer horizon radius
\be \mathcal{S}_0 =\int  \R_+ \; dx. \ee
Due to conservation of energy and charge, this integral is the same for the UBS and NUBS, and we need to examine $\mathcal{S}_1$ to witness an entropy difference between the two phases. However, it does not suffice to know the expansion of $\sqrt{g_{xx}}$ at subleading order; one also has to take into consideration the $1/n$ correction to the Killing horizon. We obtain the latter by requiring
\be g_{00} = A^{(0)}(v,x,\R_h) + \frac{A^{(1)}(v,x,\R_h)}{n} = 0, \;\;\; \text{with} \;\;\; \R_h = \R_+ + \frac{\R_h^{(1)}}{n}.\ee
Thus we find
\be \mathcal{S}_1 = \R_h^{(1)}  + \R_+ \log \left( \R_+ - \R_- \right) + \frac{\R_+}{2} G_{xx}^{(0)}(\R_+)  \ee
At equilibrium, charge density diffuses until it becomes proportional to the mass profile of the non-uniform black string, which enables us to write $q = \rho m$, with $\rho = q_0/m_0$ being the (conserved) charge-to-mass ratio of the black string. This yields
\begin{align} \mathcal{S}_1^\text{NUBS} =&\; \int \frac{1+\sqrt{1-2\rho^2}}{2} \Bigg\{ \Big( m^{\prime \prime} + m \Big) \log \left( (1+\sqrt{1-2\rho^2}) \frac{m}{2} \right)  \nonumber \\
+&\; \frac{1}{2} \left( \frac{2 \rho^2}{1-2\rho^2 + \sqrt{1-2\rho^2}} m + \frac{m^{\prime 2}}{m} \right)  \Bigg\}  dx \\
 \mathcal{S}_1^\text{UBS} =&\; \frac{2 \pi}{k} (1+\sqrt{1-2\rho^2}) \Bigg\{ \frac{m_0}{2} \log \left( (1+\sqrt{1-2\rho^2}) \frac{m_0}{2} \right)  + \frac{\rho^2}{1-2\rho^2 + \sqrt{1-2\rho^2}} \frac{m_0}{2}  \Bigg\}   \end{align}
where we have taken advantage of the fact that the coefficients multiplying $\delta m$ and $\delta q$ were constant to integrate them away. It is easy to check that the difference in the two phases' horizon area $\Delta \mathcal{S}_1 \equiv \mathcal{S}_1^\text{NUBS} -  \mathcal{S}_1^\text{UBS}$ is always positive no matter the ratio $\rho$. In particular, the entropy difference at extremality is half that of neutral strings
\be \Delta \mathcal{S}_1\vert_{\rho = \frac{1}{\sqrt{2}}} = \frac{1}{2}  \Delta \mathcal{S}_1\vert_{\rho = 0} = - \frac{2 \pi m_0}{k} \left( T_\text{NUBS}^{(1)} - T_\text{UBS}^{(1)} \right)  > 0. \label{extremaldT} \ee
\begin{figure*}[ht]
\centering
\includegraphics[width=9cm]{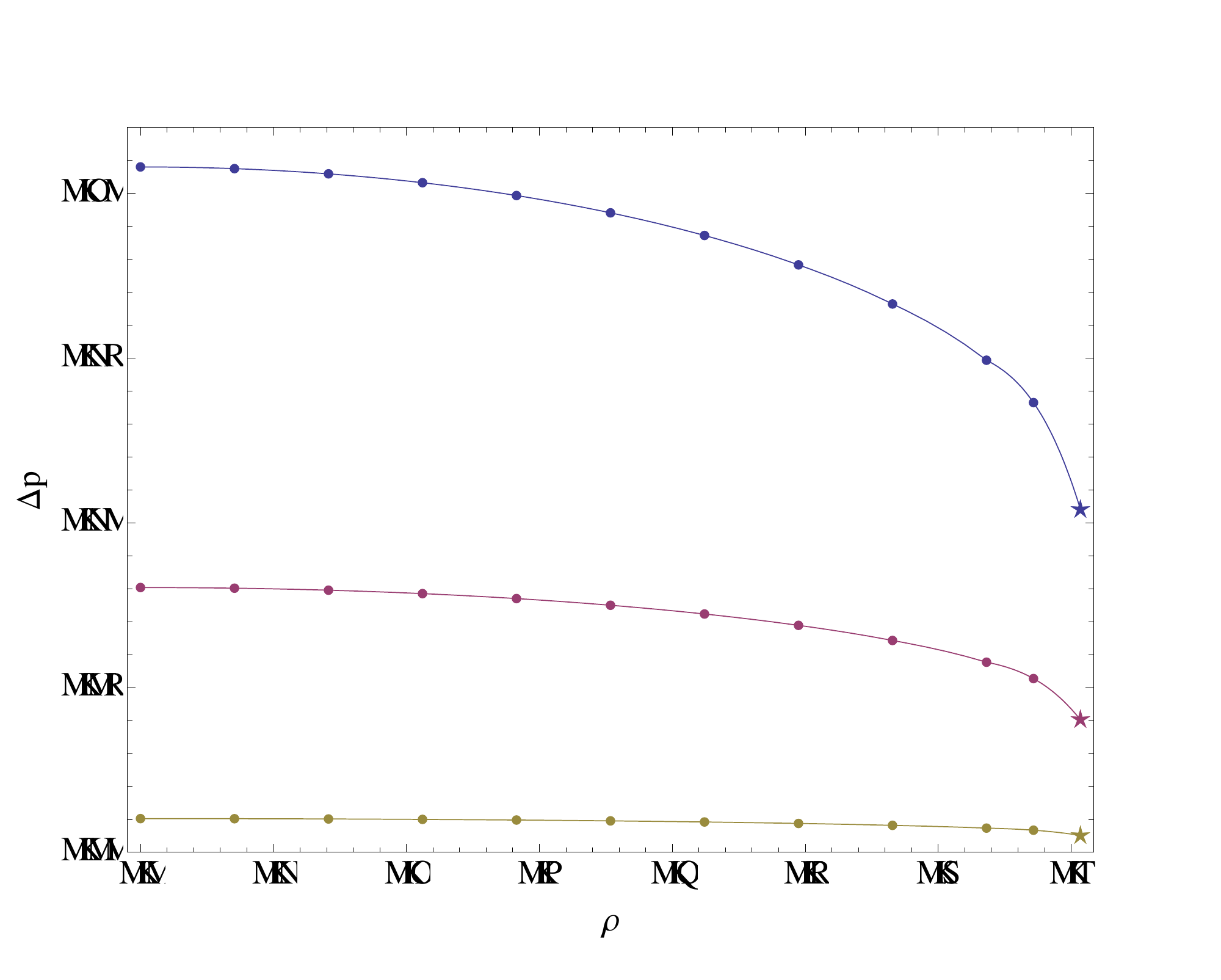}
\caption{Entropy difference per unit length as a function of the charge density for $k = \{ 0.75, 0.85, 0.95\}$, starting from the top. The numerical evolution breaks down at extremality, hence we use a star plot marker to distinguish the analytical result at $\rho = 1/\sqrt{2}$ from the others.}
\label{fig:CBSentropy}
\end{figure*}
This result indicates that the NUBS is always the preferred phase, and thus the instability persists for all charged brane configurations (as illustrated in Figure \ref{fig:CBSentropy}). However, despite the effective theory (\ref{subeqns}) admitting a smooth limit when $\rho \rightarrow \frac{1}{\sqrt{2}}$, we need to keep in mind that the large $D$ expansion formally breaks down at extremality. Nevertheless, this result corroborates the ones obtained via hydrodynamics \cite{AFRN}. This and the exact cancellation in (\ref{extremaldT}) of the pathologic divergences typically encountered at extremality both offer a positive outlook on the validity of results beyond the limits of our approximation. \\
\\
We note that the numerical results of this section and the next have been obtained by evolving small periodic perturbations around a uniform black string solution using a Runge-Kutta-Fehlberg method on a periodic Fourier grid made of 41 points. The conserved quantities $\mathcal{M}$, $\mathcal{Q}$ and $\mathcal{P}$, as well as the charge-to-mass ratio $\rho$, all remained constant during the evolution. Likewise, the integrals of $\delta m$ and $\delta p$ along the string direction were both zero to very good numerical accuracy until the final state was reached.

\subsection{Pinch-Off?}
\label{sec:pinchoff}

Let us now turn our attention back to the mass and momentum corrections $\delta m$ and $\delta p$. In principle, the contents of equations (\ref{colleqnscorr}) should provide us with a method for identifying the critical dimension $D^*$ below which the black string would \textit{pinch-off}, rather than settle on a smooth non-uniform final state. Such a transition in the nature of the final state is expected at low enough $D$. It is interesting to see how this manifests itself in the large $D$ expansion. \\
\\
Below we present criteria we impose on the solutions, and the time evolution towards those solutions, to investigate that question. While the various criteria we impose clearly indicate a tendency towards a pinch-off, we were not able to find a universal value for the critical dimension $D^*$.\\
\begin{figure*}[ht]
\centering
\includegraphics[width=10cm]{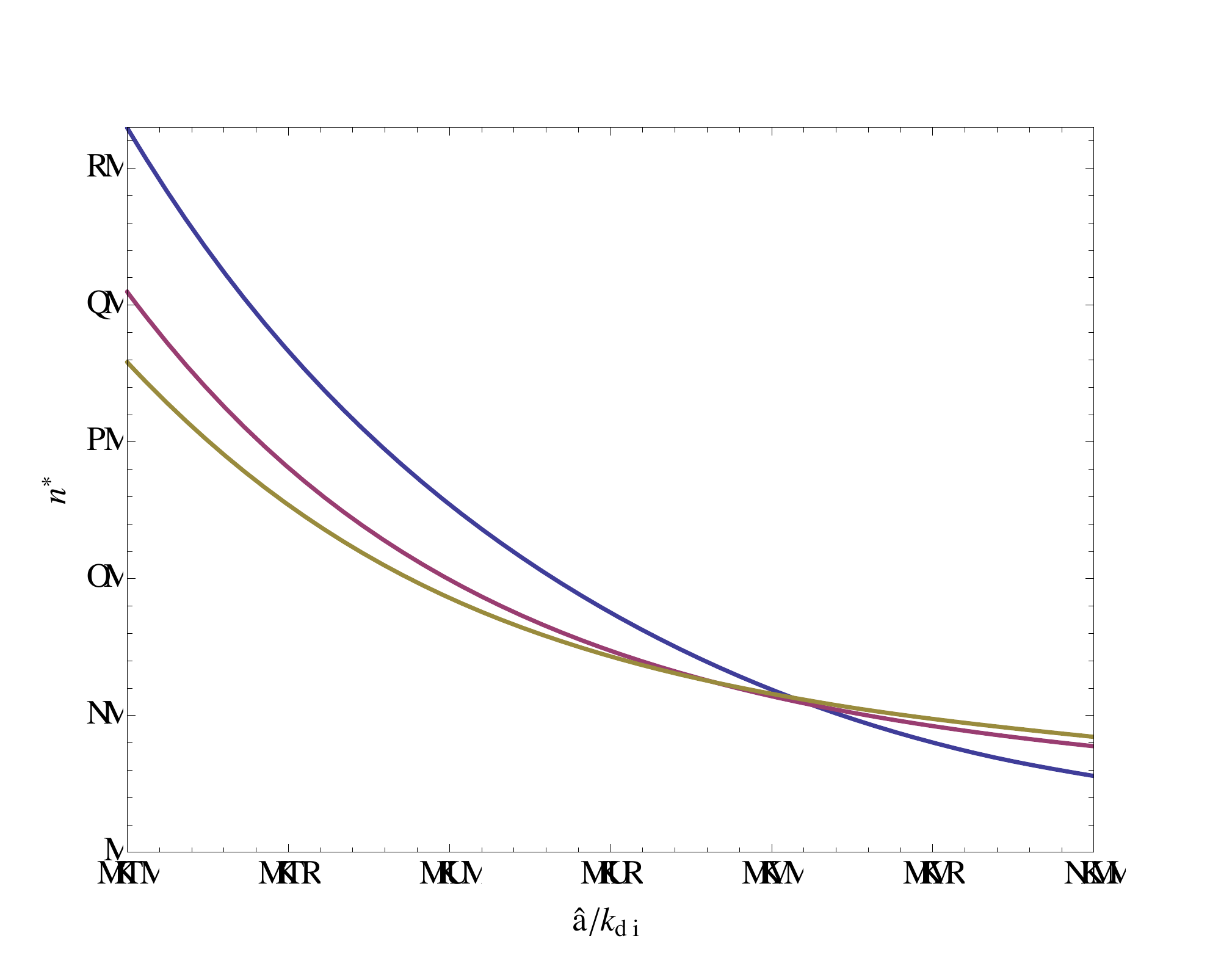}
\caption{The critical curves $n^*$ for $m_0 = \{1, 3, 5\}$, from top to bottom, obtained at NLO. Black strings with parameters above the critical curves correspond to stable NUBS, whereas we conjecture a pinch-off scenario for those below, which exhibit negative tension.}
\label{fig:CriticalTension}
\end{figure*}
\\
To determine the critical dimension $n^*$, we first define the total corrected mass density
\be M_n(x) = m(x) + \frac{\delta m(x)}{n}, \ee
and the corrected tension
\be \mathcal{T}_n = - \int \left(  \tau_{xx}^{(0)} + \frac{ \tau_{xx}^{(1)} }{n} \right) dx, \label{corrtension} \ee
which are both gauge invariant quantities. \\
\\
One attempt at diagnosing the stability of the black string at subleading order using our knowledge of $m(x)$ and $\delta m(x)$ is to find $n^*$ such that $M_{n*}(x)$ becomes locally negative. One can do so either by looking at the dynamical evolution, or by examining the properties of the final state only. It turns out that the dynamical evolution of the collective variables is highly sensitive to the size of the initial perturbations around the uniform solution $m(x) = m_0$. As such, this method does not provide a reliable stability diagnostic.  As for the shape of the end-point of the dynamical evolution, the final shape of $\delta m(x)$ does depends solely on the static profile $m(x)$. This makes it possible to find $n^*$ such that $M_{n^*}(x) < 0$ locally, but this method has not yielded accurate results. \\
\\
An alternate, more successful, method to identify the critical dimension $n^*$ uses the corrected tension (\ref{corrtension}). Indeed, it is possible to find a critical curve $n^*$ as a function of the dimensionless wavenumber $k/k_{GL}$ by assuming that the fate of a NUBS with negative tension is to pinch-off. Our results are summarized Figure \ref{fig:CriticalTension} for three different initial configurations $m_0$. Note that as we vary $m_0$ we change $k_\text{GL}$, so this is an alternate way to scan the the ``thickness" $k/k_\text{GL}$ . However, while the qualitative features are similar, we still see slight differences between the three curves,  which are either an artifact of early truncation in the $1/n$ expansion or a sign that negative brane tension is a sufficient but not necessary condition for pinching-off. Indeed, nothing stops a pinch-off from happening at positive tension, and as such our critical curves may be thought of as an approximate probe until further investigation. \\
\\
While the results we obtain preclude us from assigning  an unambiguous value for the critical dimension, it can serve as a bound on the dimension in which the final state pinches off, and it illustrates the dependence of the critical dimension on the brane thickness. It is interesting to note that the dimensions near the critical point $k = k_\text{GL}$ are quite close to the expected value $n^* = 9$.

\section{Two Dimensional Non-Uniform Phases}
\label{sec:lattice}

\begin{figure*}[ht]
\begin{subfigure}{0.5\textwidth}
 \includegraphics[width=8.4cm]{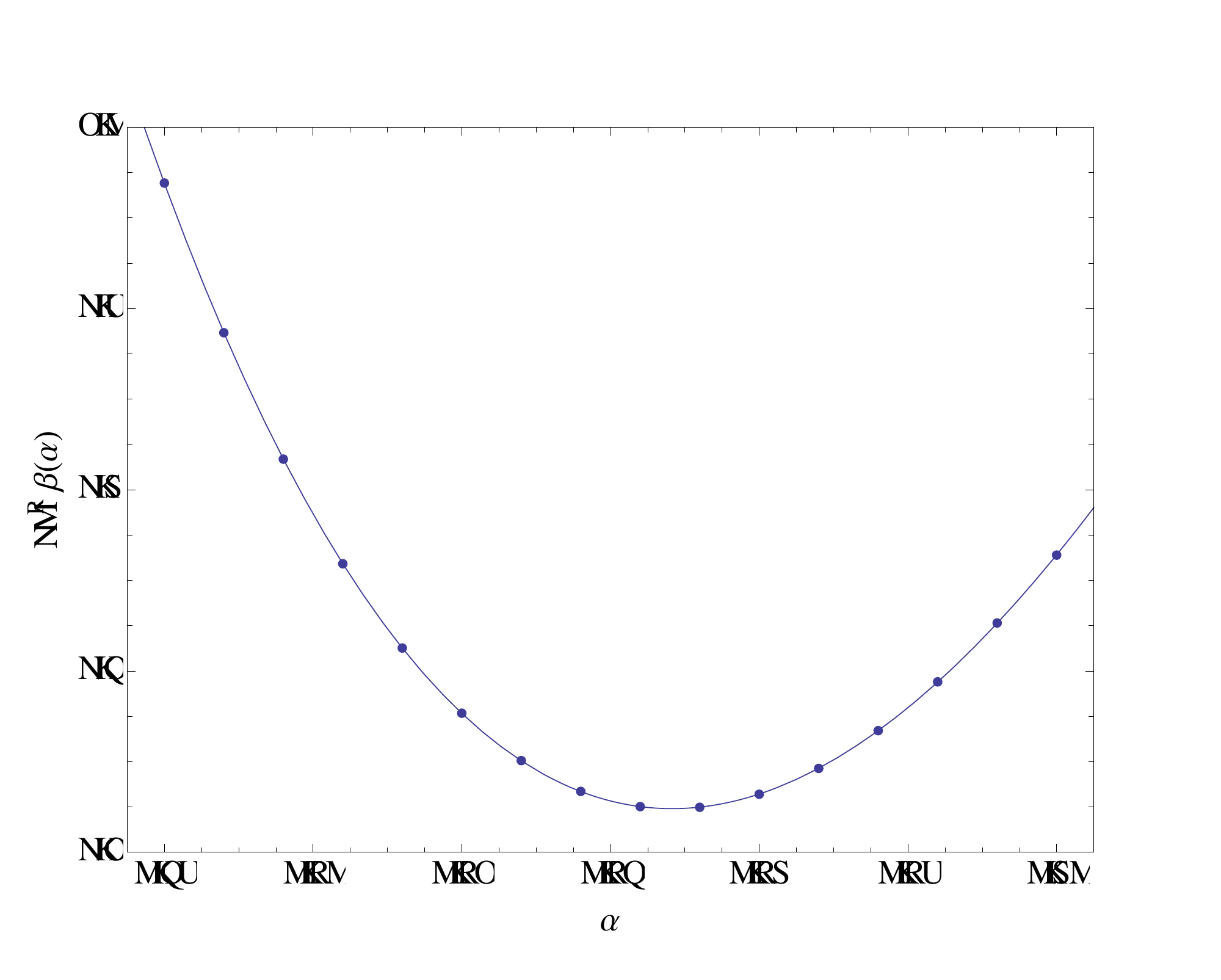}
  \subcaption{The tension as a function of the opening angle for $k=0.6$. The minimum is located at $\alpha \approx 0.548$. }
\label{fig_alpha_k=0.6}
\end{subfigure}
\hfill
\begin{subfigure}{0.5\textwidth}
    \includegraphics[width=8.4cm]{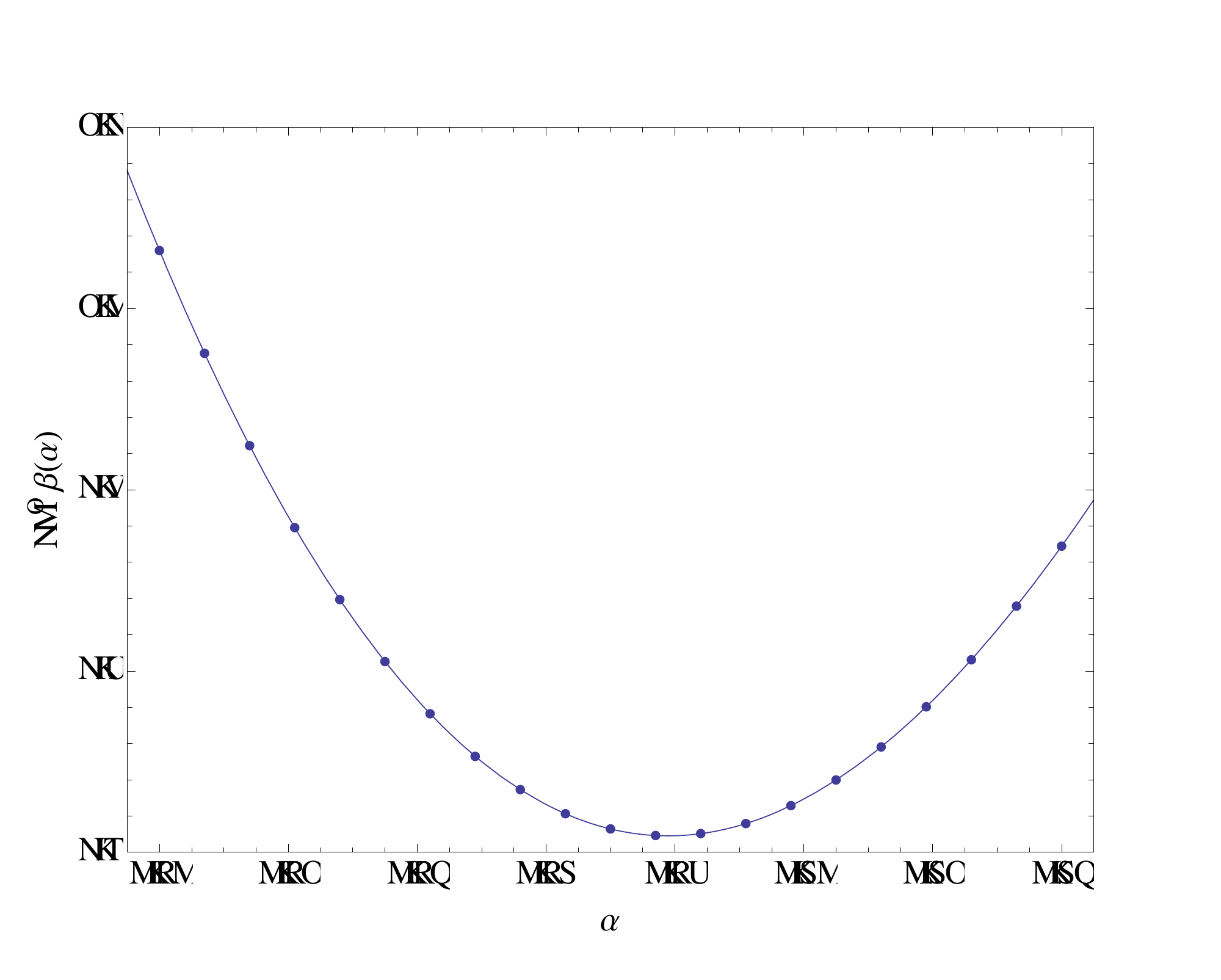}
\subcaption{The tension as a function of the opening angle for $k=0.8$. The minimum is located at $\alpha \approx 0.579$.}
\label{fig_alpha_k=0.8}
\end{subfigure}
\caption{The enthalpy of the Bravais lattices reaches its minimum closer to $\alpha = \pi/6 \approx 0.524$ as the size of the cell increases. Note that the numerical evolution becomes unstable for large $\alpha$ and large $L$, thus preventing us from probing larger oblique cells.}
\label{fig:alphaplots}
\end{figure*}

We now move to discuss the case of unstable membranes, for which there are two independent modes of Gregory-Laflamme instabilities.   Similar to the one-dimensional case we find that in the leading order in the large $D$ limit, the final state is a smooth and non-uniform configuration which we call a lattice. Using the tools developed above, we  study the phase diagram and determine the preferred size and shape of this lattice configuration. \\
\begin{figure*}[h!]
\begin{subfigure}{0.5\textwidth}
 \includegraphics[width=8.1cm]{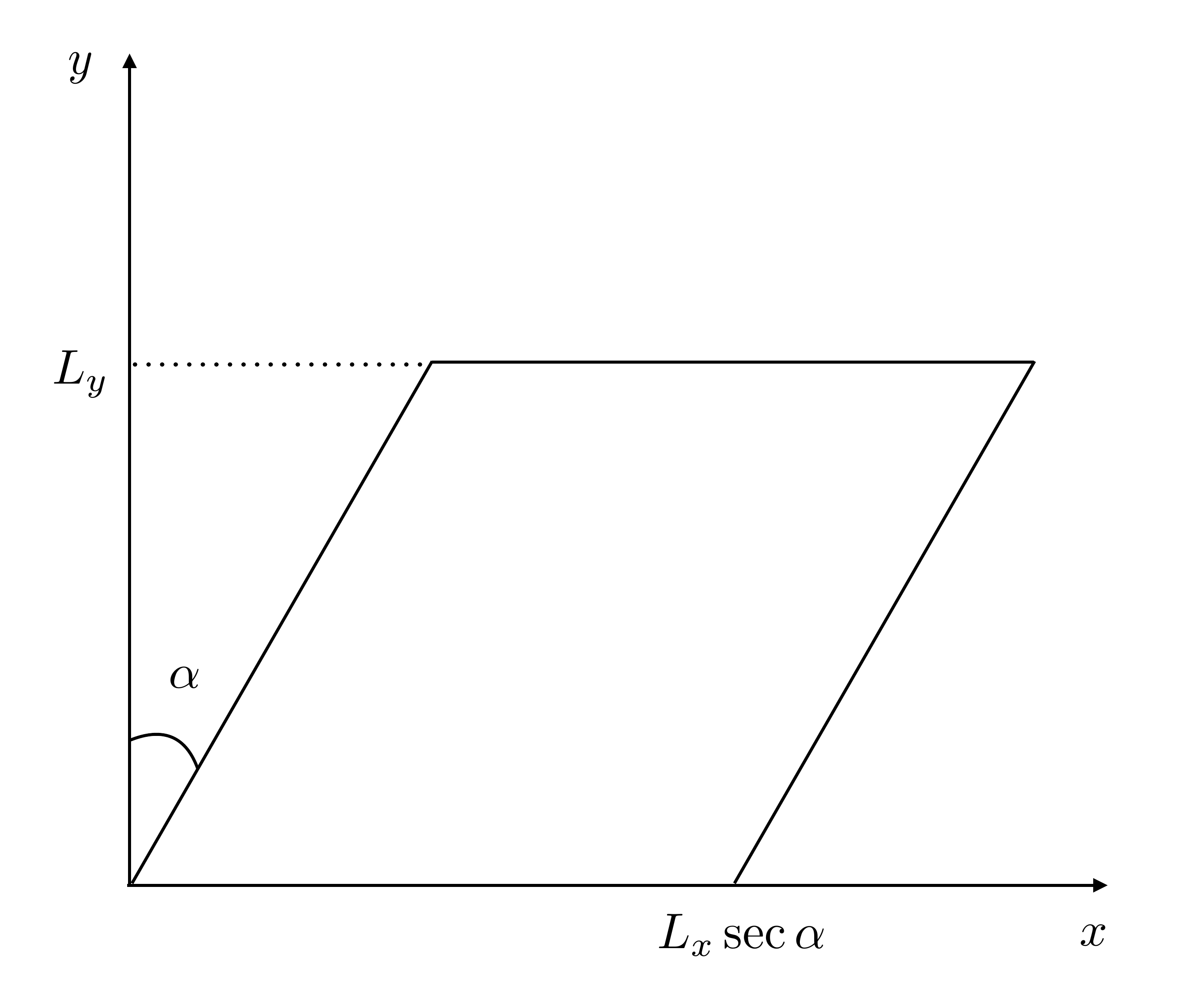}
  \subcaption{Lattice cell in $(x,y)$ coordinates}
\label{fig_latticexy}
\end{subfigure}
\hfill
\begin{subfigure}{0.5\textwidth}
    \includegraphics[width=8.1cm]{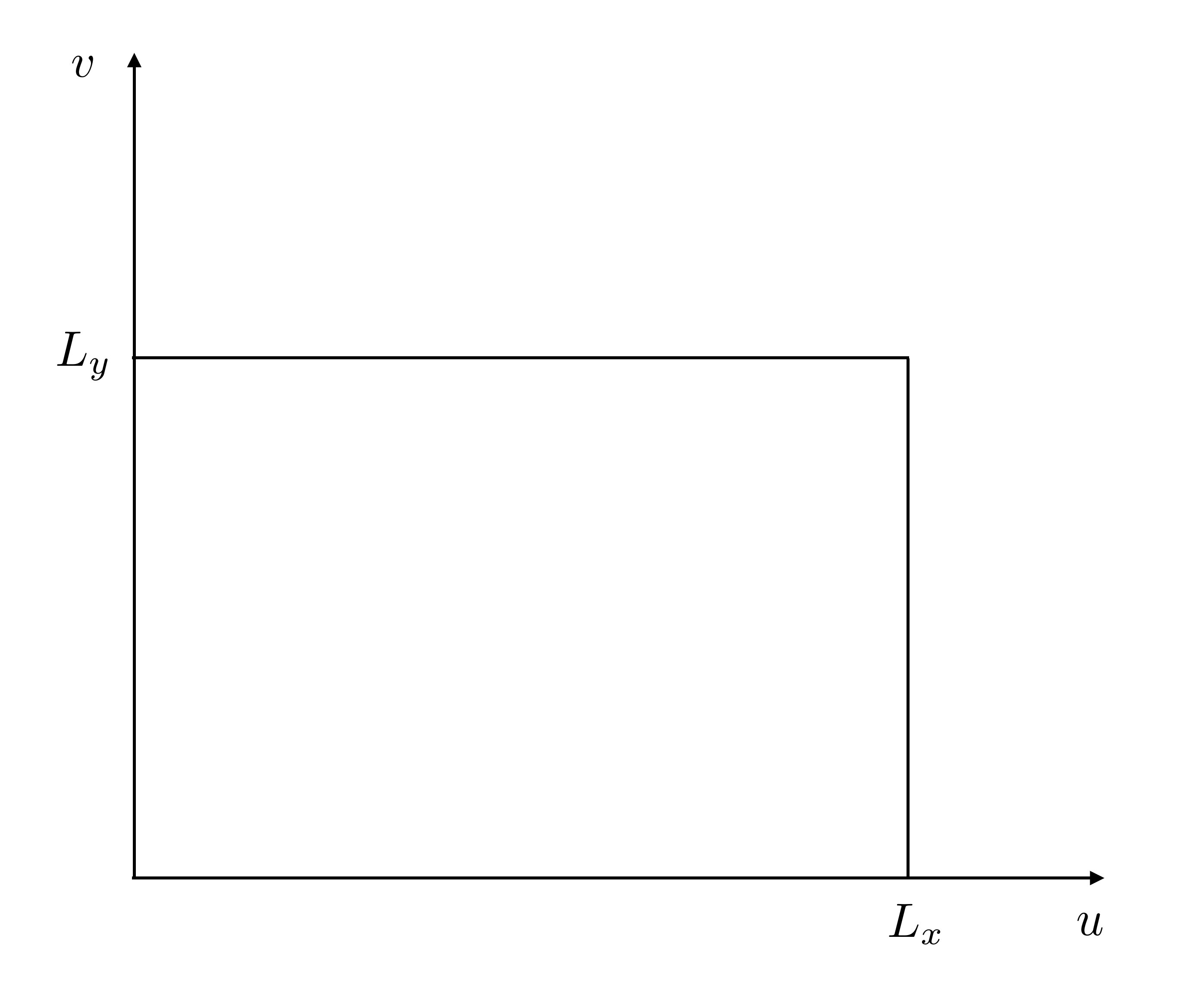}
\subcaption{Lattice cell in $(u,v)$ coordinates}
\label{fig_latticeuv}
\end{subfigure}
\caption{The change of coordinates from $(x,y)$ to $(u,v)$ maps oblique lattice cells (a) to rectangular ones (b). Their area is $\mathcal{A}_\text{cell} = L_x L_y \sec \alpha$. }
\label{fig:lattices}
\end{figure*}
\\
In two dimensions, it is possible to construct periodic black brane configurations over oblique lattices. The lattices can be described by two  vectors, describing the periodicities of the system:
\be \mathbf{k}_x = \frac{2 \pi}{L_x} (\cos \alpha, - \sin \alpha), \;\;\; \mathbf{k}_y = \frac{2 \pi}{L_y} (0,1), \;\;\;\;\; \text{with} \;\;\; 0 \leq \alpha \leq \pi/2. \ee
Thus we parametrize possible non-uniform solutions  solutions by the three parameters $(L_x, L_y, \alpha)$. Furthermore, since we are mostly interested in the preferred shape of the non-uniform configuration, we take $L_x = L_y = L$. The angle $\alpha$ characterizes then the shape: special cases include $\alpha = 0$ for checkerboard lattices, $\alpha = \pi/6$ for triangular lattices, and $\alpha = \pi/2$ for stripes. \\
\\
For the purpose of constructing the solutions, it is easier to work with the coordinates $(u,v)$ defined by
\be u = x \cos \alpha - y  \sin \alpha, \;\;\; v = y. \label{varchange} \ee
\noindent In these coordinates (illustrated in Figure \ref{fig:lattices}), periodic boundary conditions are simply
\be (u,v) \equiv (u + L_x n_x,  v + L_y n_y). \ee
for any integers $n_x,n_y$.\\
\\
In order to make meaningful comparison between lattices of different size and shape, we need to work with the right thermodynamic potential. Instead of fixing the size and the shape of the unit cell, we instead fix the conjugate variables: the tensions in different directions.  See \cite{Donos:2013cka} for a general discussion, and \cite{Donos:2015eew} for a recent application closely related to the current discussion.\\
\\
The first law of black brane dynamics, in the micro-canonical ensemble, can be written as
\be dM = \kappa dA + \mathcal{T}^{ab} dV_{ab}, \ee
where $\mathcal{T}^{ab}$ are related to the tensions along brane directions, and $V_{ab}$ is a matrix of periodicities. To fix the conjugate variables instead of the size and shape of the brane configuration, we define the ``enthalpy" $H$ of the brane\footnote{The conventional entalphy is obtained by Legendre transform with respect to the total volume, to work with fixed pressure instead.} as the Legendre transform
\be H \equiv M - \mathcal{T}^{ab} V_{ab}. \ee
Our goal is to minimize this new potential. But first, we need to find an expression for the tensions $\mathcal{T}^{ab}$. These are usually obtained from the quasilocal stress tensor at $\R \rightarrow \infty$, which we have already found in (\ref{Txx0}):
\be \tau^{ab} = \partial_a \partial_b m - m \delta_{ab} + \frac{p_a p_b}{m} - \left(\partial_a p_b +\partial_b p_a \right). \ee
We identify this boundary stress tensor as the source for the tensions, such that 
\be \mathcal{T}^{ab} = \langle \tau^{ab} \rangle = \iint_\text{cell} \tau^{ab} dx dy\Big/\iint_\text{cell} dx dy =  \frac{1}{L^2}  \iint_\text{cell} \tau^{ab} du dv. \ee
In the orthogonal coordinates $(u,v)$, only the pressures $\mathcal{T}^{uu}$ and $\mathcal{T}^{vv}$ should contribute, and as such we take $V_{mn} = L^2 \delta_{mn}$ for $m,n = u,v$. Since we are working at constant mass, the quantity we must minimize is the tension $\beta$ given by
\be \beta(\alpha)= - \mathcal{T}^{mn} \delta_{mn} = -\frac{1}{L^2} \iint_\text{cell} \left( \cos^2 \alpha \; \tau^{xx} - 2 \cos \alpha \sin \alpha \;  \tau^{xy} + (1+\sin^2 \alpha) \; \tau^{yy} \right) du dv \label{shear}. \ee
As expected, the enthalpy of oblique lattices includes contributions from the shear components of the stress tensor $\tau^{xy}$. \\
\\
It is straightforward to apply the change of variables (\ref{varchange}) to (\ref{subeqns}) in order to find the inhomogeneous solutions on the oblique lattice numerically. For that purpose we discretize the lattices on a $31 \times 31$ periodic Fourier spectral grid, and we used the fifth order Runge-Kutta-Fehlberg time-stepping algorithm to perform the time evolution towards the stable inhomogeneous solution. \\
\\
Once we obtain the solutions for different shapes, we can find which shape is preferred --- our results are illustrated in Figure \ref{fig:alphaplots} for two different choices of lattice size. We find that the minimum of $H$ is reached for lattices with opening angles close to $\alpha = \pi/6$, which corresponds to the triangular lattice. Unsurprisingly, the position of the minimum depends on the size of the cell. Based on these results, we expect that for asymptotically large lattices the triangular lattice is the preferred configuration, and the slight deviations we see are due to finite size effects.\\
\\
One can repeat the exercise with respect to the size of the preferred configuration. Indeed, in the one-dimensional case, where there is a size but not shape parameter, the tension $\beta$ of a black string with mass density $m_0$ decays exponentially with the string length $L$
\be \beta(L) = -\langle \tau^{xx} \rangle \sim m_0 \; e^{-a(L-2\pi) }, \;\;\;\;\; \text{with} \;\;\; a \approx 1.827, \ee
meaning that the size of the preferred configuration is asymptotically large.

\section{Conclusion}
\label{sec:conclusion}

The principal focus of our work was to determine the fate of extended black objects,  in the approximation where the number of dimensions is large. The tools of general relativity at large $D$ have proven useful at unveiling robust properties of higher dimensional black holes to surprising accuracy, and our hope is that likewise the results presented in this paper hold up beyond the asymptotic limit of that approximation. \\
\\
The main loose end left in this work is determining whether negative brane tension is an appropriate test to accurately determine the fate of the black string instability.  While we see indications that the pinch-off scenario is likely as the final state at sufficiently low $D$, as well as a non-trivial dependence of the associated critical dimension on the string thickness, we have not obtained a precise unambiguous result nor succeeded in reconciling these features of our solution with the current results in the literature. We hope to return to this in the future. \\
\\
It would also be an interesting endeavour to explore the dynamics of charged dilatonic Kaluza-Klein black holes given the existence of exact uniform solutions. Such a direct comparison would expand our understanding of the effects of charge on the stability of black strings in more general scenarios.

\section*{Acknowledgements}
\label{sec:acknowledgements}

This work is supported by a Discovery grant from the Natural Sciences and Engineering Research Council of Canada. We thank Roberto Emparan and Mukund Rangamani for interesting conversations on the large $D$ limit, and to Kristan Jensen for collaboration in the initial stages of this project.
\bigskip

\bibliographystyle{jhep}
\bibliography{chargedstring}

\end{document}